\documentclass[a4paper,11pt]{article}
\usepackage{jcappub} 
\usepackage{lineno}
\usepackage{float}
\usepackage{multirow}

\arxivnumber{2405.06750} 
\title{Scalar Field Dominated Cosmology with Woods-Saxon Like Potential}

\author[a]{Sreerag Radhakrishnan}\emailAdd{sreeragradhakrishnan01@gmail.com}

\author[b]{Sarath Nelleri}\emailAdd{sarathn@iitk.ac.in}
\author[b]{Navaneeth Poonthottathil}\emailAdd{navaneeth@iitk.ac.in}
\affiliation[a]{Department of Physics, Government Brennen College,\\Dharmadam, Thalassery, 670106, Kerala, India}
\affiliation[b]{Department of Physics, Indian Institute of Technology,\\Kanpur, 208016, India}

\abstract{Dark energy can be characterized by a canonical scalar field, known as quintessence. Quintessence allows for a dynamical equation of state $-1 \le \omega \le -\frac{1}{3}$. A previous study by Oikonomou and Chatzarakis have shown that a scalar field model with a Woods-Saxon like potential can successfully explain the early inflation. In this work, we consider a quintessence model with a potential of similar form to explain the late time acceleration. The model is studied at late phase assuming flat cosmology, and the model parameters are constrained using Type Ia supernova data and Observational Hubble data. In particular we employ Markov Chain Monte Carlo methods for the Bayesian inference of these parameters. We obtain the value of the Hubble constant $H_0 \sim 68 \text{ km s}^{-1} \text{Mpc}^{-1}$ and the matter energy density parameter $\Omega_{m_0} \sim 0.30 $, which are in close agreement with the values obtained from the Planck CMB data, assuming the $\Lambda$CDM model. Computation of the $\chi^2_{min}$, AIC and BIC reveal that this model is slightly preferred according to AIC and $\chi^2_{min}$ criteria, while the $\Lambda$CDM is preferred according to BIC. We demonstrate that the model possesses a stable attractor in the asymptotic future, which confirms the dynamical stability of the model. Thus, this model may be considered as a potential alternative to the $\Lambda$CDM.}

\begin{document}
\maketitle
\flushbottom

\section{Introduction}
Contemporary cosmology attempts to unmask the mystery of dark energy, which drives the (observed) late-time accelerated expansion of the Universe. The initial evidence for this was from the observations of Type Ia Supernovae (SNeIa) as presented in Ref. \cite{agreiss1998,perlmutter1999}, which confirmed the acceleration of the Universe. According to such observations, the Universe is assumed to be dominated by a mysterious component called dark energy, constituting about $70\%$ of the present Universe, the rest consisting of matter, of which normal baryonic matter is only around $4\%$, and the remaining is in the form of "invisible" dark matter. This inference has also been confirmed by other observations such that of the CMB \cite{tegmark1999,spergel2003,tauber2014planck,planck2} and Baryonic Acoustic Osillations \cite{eisenstein2005,sollerman2009,tegmark2004cosmological}.

Dark energy is characterized by a perfect fluid with a negative equation of state, $\omega = P/\rho$. One of the simplest models for dark energy is the cosmological constant $\Lambda$\cite{RevModPhys.75.559,peebles1993principles}. The cosmological constant has a characteristic equation of state $\omega = -1$. This is the currently accepted model of dark energy, by the standard $\Lambda$CDM. Physical explanation of the cosmological constant may be attributed to the zero-point quantum vacuum fluctuations in the Universe \cite{sahnilambda}, given by $\Lambda = 8\pi G\rho_{vac}/c^2$. However, theoretical predictions of $\rho_{vac}$ do not match with the current observations, by several orders of magnitude \cite{weinberg1989,quartin2008}. This cosmological constant problem, suggests that the mentioned explanation for the cosmological constant is unsatisfactory. In addition to this, $\Lambda$CDM encounters problems such as the Cosmological Coincidence Problem \cite{veltenccp} and the tensions in the inferred values of various parameters between low redshift data \cite{agreisstension2,agreisstension1,joudaki2020} and high redshift data \cite{lusso2019,refId0}. Attempts to resolve these issues have been ongoing and many possible solutions have been suggested in the form of alternate cosmological models. These include the modified gravity models (Ref. \cite{Odintsov_2021,Clifton_2012,koyama2016,silvestri2013,bamba2015}), which modify the curvature term in Einstein Field Equations, and modified (or dynamical) dark energy models which modify the form of the energy-momentum tensor in the field equations \cite{dynde,pan2019,karwal2022,murgia2021,krishnan2021,bamba2012,niedermann2020,alam2004,peracaula2021,antoiadis2007}.

Such kinds of modified dark energy models characterize dark energy as a specific form of matter, such as quintessence \cite{caldwell1998}, k-essence \cite{picon2000} or the Chaplygin gas \cite{KAMENSHCHIK2001265}. The quintessence is a particular model in which a scalar field is slow rolling down a potential that is similar to that of early Universe inflation models \cite{zlatev1999}. Reviews of various quintessence models can be found in Ref. \cite{tsujikawa2013,caldwell2005,chiba2002,brax2000,jesus2008,perrotta1999,chimento2000,affleck1985,nomura2000,kim2003,panda2011,carollquint}. Another kind of scalar field model is k-essence, which, unlike quintessence, does not have a canonical kinetic term. These models have also been studied extensively \cite{picon2001,malquarti2003,rendall2006,piconkessence} much like quintessence. The advantage of these scalar field models is that some of these models can address the fine tuning and coincidence problem, all the while explaining the late-time cosmic acceleration. These models are distinguished by their dynamical equation of state $\omega$, which is not the case for $\Lambda$CDM.

The evolution of $\omega$ is the key factor in distinguishing quintessence models. Quintessence can be of two broad types : (i) thawing models and (ii) freezing models. In thawing models, at early times, the field was frozen by Hubble damping at a value different from its minimum potential, started to roll down to the minimum at late times. These models have an early equation of state $\omega_\phi \approx -1$ and becomes less negative over time, that is, $\omega_\phi' > 0$. In the case of freezing models, the scalar field is already rolling towards its minimum potential, but slows down at later times when it dominates the Universe. This means that the equation of state "freezes" close to the present time and $\omega_\phi ' < 0$. 

Quintessence models, as mentioned, have been explored as alternatives to address the shortcomings of the standard $\Lambda$CDM. Various quintessence models have been studied in literature, giving specific forms to the potential energy of the scalar field\cite{PhysRevD.57.4686,Alho_2015,hill1988,amendola2000,amendola2014,LaurJärv_2004,PhysRevD.80.123521,Fang_2009}. But it is important that the model exhibits a stable attractor solution in the asymptotic future, so that a large assortment of initial conditions can lead to the same stable future. Else it would have the same fine tuning problem as the concordance model. In addition, the considered potential must be able to reproduce the cosmic acceleration. Dynamical system analysis is necessary to verify the existence of such a stable attractor. This technique is widely used to cosmology to analyse the stability of a model\cite{coley2013dynamical}, especially for scalar field models\cite{BOHMER201211}. The stability of critical (or "fixed") points are examined mainly using the linear stability theorem,\cite{raushan2021linear} and to analyse non-hyperbolic critical points, centre manifold theorem\cite{wigginsintroduction,carr2012applications} is employed.

The aim of this paper is to construct a quintessence model with a potential which has a similar form as that of the Woods-Saxon potential of nuclear physics. It is essential to emphasize that while we draw inspiration from the form of the Woods-Saxon potential, this model is entirely distinct in its interpretation within the context of quintessence cosmology. A similar scalar field approach with the Woods-Saxon potential has shown promising results as an inflationary model for the primordial era of the Universe. The authors of \cite{woods-saxoninflation} were able to successfully realize an inflationary phase along with a consistent reheating epoch with the scalar field model, which is compatible with the Planck CMB data. This study aims to test the phenomenological implications in the late-time accelerated expansion of the Universe with a Woods-Saxon like potential in a quintessence scenario. We constrain the model with observations, and study the evolution of the Universe in this model. In our dynamical system analysis, we obtain one non-hyperbolic point which had to be studied using the centre manifold approach. This approach makes it possible to study the dynamics of the system near the critical point by reducing its dimensionality. The analysis reveals the existence of a stable attractor solution, indicating that the Universe is dynamically stable within the framework of this model.

This paper is organized as follows:
In Sec. \ref{Model}, we present our proposed model, which will be referred to as the Woods-Saxon Quintessence or WSQ hereafter. In Sec. \ref{Dataanalysis}, we employ Bayesian inference to constrain the model parameters using observational data, specifically the Type Ia supernova data and observational Hubble parameter data. In Sec. \ref{evolution}, we investigate the evolution of various cosmological parameters within the framework of the WSQ model. In Sec. \ref{dynamicalanalysis}, we test the dynamical stability of the proposed WSQ model by performing dynamical system analysis. Finally, we conclude our work in Sec. \ref{conclusion}.

\section{Woods-Saxon Quintessence}\label{Model}
We consider a quintessence in the presence of non-relativistic matter, represented by a barotropic perfect fluid. As we are focusing on the late-time cosmology, the effect of radiation component is negligibly small and hence, we only consider the contribution from matter and field. The action would be given by

\begin{equation}\label{actionquint}
    S = \int d^4x \sqrt{-g} \left[\frac{1}{2}\frac{R}{\kappa^2} - \frac{1}{2}g^{\mu\nu}\partial_\mu\phi\partial_\nu\phi - V(\phi) \right] + S_m,
\end{equation}
where $\kappa^2 = 8\pi G$, $S_m$ is the matter action, $R$ is the Ricci scalar, $g^{\mu\nu}$ and $g$ are the metric and its determinant respectively. Here, $V(\phi)$ is the potential over which the field $\phi$ is rolling towards some minima. We study the quintessence in the flat Friedmann-Lemaitre-Robertson-Walker (FLRW) background with the curvature parameter $k = 0$.
For a quintessence field $\phi$, the pressure and energy density are given by $P_\phi = \frac{\dot{\phi}^2}{2} - V(\phi)$ and $\rho_\phi = \frac{\dot{\phi}^2}{2} + V(\phi)$ respectively, where the overdot represents derivative with respect to cosmic time $t$. Then, the equation of state takes the form
\begin{equation}\label{equationofstate}
    \omega_\phi = \frac{P_\phi}{\rho_\phi} = \frac{\frac{\dot{\phi}^2}{2} - V(\phi)}{\frac{\dot{\phi}^2}{2} + V(\phi)}.
\end{equation}
Here, we assume that the dark matter and field are non-interacting and hence satisfies individual conservation equations
\begin{align}
   \label{fieldcontinuity} \dot{\rho_\phi} + 3H(\rho_\phi + P_\phi) = 0,\\
    \dot{\rho_m} + 3H\rho_m = 0,
\end{align}
where the dark matter is assumed to be pressureless, giving an equation of state $\omega_m = 0$.
Introducing the forms of $\rho_\phi$ and $P_\phi$ in Eq. (\ref{fieldcontinuity}), we obtain the Klein-Gordon equation,
\begin{equation}\label{continuity}
    \ddot{\phi} + 3H\dot{\phi} + V_{,\phi} = 0,
\end{equation}
where $ V_{,\phi} = \frac{dV}{d\phi}$ and $H = \frac{\dot{a}}{a}$ is the Hubble parameter.
The Friedmann equations in this framework can be obtained as,
\begin{align}
    &\label{mot1}3H^2 = \kappa^2\left[ \frac{\dot{\phi}^2}{2} + V(\phi) + \rho_m \right],\\
    &\label{mot2}2\dot{H} = -\kappa^2[\dot{\phi}^2 + \rho_m].
\end{align}
We consider a potential $V(\phi)$ which resembles the Woods-Saxon potential,
\begin{equation}\label{woodspotential}
    V(\phi) = \frac{2V_0}{e^{\kappa (\phi - \phi_0)} + 1},
\end{equation}
where $V_0$ and $\phi_0$ are the values of the potential energy and scalar field at present time, i.e., at redshift $z = 0$, respectively.
Analytical solution to Eq. (\ref{mot2}) is challenging with the potential $V(\phi)$. In order to solve Eq. (\ref{mot2}) numerically, we introduce three dimensionless parameters,
\begin{equation}\label{parameters}
    x = \frac{\dot{\phi}\kappa}{\sqrt{6}H_0},
    \hspace{0.2in} y = \frac{\sqrt{V(\phi)}\kappa}{\sqrt{3}H_0},\hspace{0.2in} h = \frac{H}{H_0},
\end{equation}
where $H_0$ is the Hubble parameter at present.
Taking the derivatives of $x$, $y$ and $h$ with $z$ and using Eq. (\ref{continuity}), (\ref{mot1}) and (\ref{mot2}), we obtain three coupled differential equations,
\begin{align}
    \label{model1}&\frac{dx}{dz} = \frac{3x}{1+z} - \sqrt{\frac{3}{2}}\frac{y^2}{h(1+z)}\left[1 - \frac{1}{2}\left(\frac{y}{y_0}\right)^2\right],\\
    \label{model2}&\frac{dy}{dz} = \sqrt{\frac{3}{2}} \frac{xy}{h(1+z)}\left[1 - \frac{1}{2}\left(\frac{y}{y_0}\right)^2\right],\\
    \label{model3}&\frac{dh}{dz} = \frac{3}{2} \frac{h^2 + x^2 - y^2}{h(1+z)}.
\end{align}
Here we have defined $y_0 = y\,(z = 0)$. On solving these equations numerically, we obtain $x$, $y$, and $h$ as a function of $z$. These variables are related through the equation,
\begin{equation}\label{constraint1}
    h^2 = x^2 + y^2 + \Omega_m,
\end{equation}
where we have defined $\Omega_m = \rho_m/\rho_{c_0} = \rho_m\kappa^2/3H_0^2$, where $\rho_{c_0} = 3H_0^2/\kappa^2$ is the present value of the critical density.
We define the energy density parameter of the field , $\Omega_\phi$ using Eq. (\ref{parameters}) in terms of the variables $x$ and $y$ as,
\begin{equation}
    \Omega_\phi = \frac{\kappa^2\rho_\phi}{3H_0^2} = x^2 + y^2.
\end{equation}
The equation of state can be written using Eq. (\ref{equationofstate}) as,
\begin{equation}
    \omega_\phi = \frac{x^2 - y^2}{x^2 + y^2}.
\end{equation}
In this model, $H_0$, $\Omega_{m_0}$ and $\Omega_{\phi_0}$ are the physical quantities of interest that are to be determined. Our aim is to estimate these parameters utilizing the latest observational probes.
\section{Data Analysis}\label{Dataanalysis}
In order to perform the data analysis, we have to solve the coupled differential equations presented in Eq. (\ref{model1})-(\ref{model3}), which require the initial conditions $x_0$, $y_0$ and $h_0$. These parameters are constrained by the equation $1 = x_0^2 + y_0^2 + \Omega_{m_0}$, where $h_0 = 1$. Thus, the WSQ model has effectively three independent free parameters $H_0$, $\Omega_{m_0}$ and $y_0$. In order to constrain these parameters, we employ Markov Chain Monte Carlo methods for Bayesian inference \cite{bayesinthesky}. The primary input for the Bayesian inference is the prior range for each model parameter. We assume a sufficiently large prior range for $H_0$ between $40$ and $100$, and $0$ to $1$ for $\Omega_{m_0}$. The prior range for $y_0$ is assumed to be $0$ to $1$ as the present density parameters of the components add to unity, and $y$ must be positive by definition. The parameters are constrained using Observational Hubble Data (OHD) and Type Ia Supernova (SNIa) data. The dataset we used for the OHD data is from Ref. \cite{OHDset}. We considered two different datasets for the SNIa data - The Pantheon+ dataset\cite{Pantheonplus2} and the five year Dark Energy Survey (DES) data \cite{abbott2024dark,vincenzi2024dark}.
The OHD dataset consists of $77$ redshift vs Hubble parameter values in the redshift range $0 \le z \le 2.36$, obtained from cosmic chronometers, BAO signal in galaxy distribution and BAO signals in the $Ly\alpha$ forest distribution. The Pantheon+ data comprises the sample of $1701$ Type Ia supernova light curves of 1550 distinct supernovae discovered in the redshift range $0 \le z \le 2.3$ \cite{Pantheonplus1}. We will refer to the combination of Pantheon+ and OHD dataset as Pantheon+OHD. The DES data consists of a total sample of $1829$ comprising of $1635$ DES SNe and $194$ external low-redshift samples in the redshift range $0.025 \le z \le 1.13$. We will refer to the combination of DES and OHD datasets as the DES+OHD. The SNIa datasets consist of redshift $z$ against apparent magnitude ($m$) obtained from their respective surveys.
The apparent magnitude of SNIa, $m$ is obtained from the expression,
\begin{equation}
    m(z) = 5\,log_{10}\left[\frac{d_L(z)}{Mpc}\right] + M + 25,
\end{equation}
where $M$ is the absolute magnitude of SNIa and $d_L$ is the luminosity distance and for flat cosmology ($\Omega_k = 0$), it is given by,
\begin{equation}
    d_L(z) = c(1+z) \int_0^z \frac{dz'}{H(z')},
\end{equation}
where $c$ is the speed of light in vaccuum expressed in km/s.
The $\chi^2$ value for OHD data is computed as,
\begin{equation}
    \chi^2_{OHD} = \sum_{i = 1}^{77} \left[\frac{H_{obs}(z_i) - H_{th}(z_i,H_0,\Omega_{m_0},y_0)}{\sigma(z_i)}\right]^2,
\end{equation}
where  $H_{obs}$ and $H_{th}$ are the values of Hubble parameter observed and that computed using the WSQ model respectively, and $\sigma$ is the error in the measurement.
The $\chi^2$ for the SNIa Pantheon+ data is calculated using the expression,
\begin{equation}
    \chi^2_{SNIa} = \Delta \vec{D}^T C^{-1}_{stat+syst} \Delta \vec{D},
\end{equation}
where $\Delta \vec{D}$ is the vector of distance modulus residual given by,
\begin{equation}
    \Delta \vec{D} = \mu(z_i) - \mu_{th}(z_i,H_0,\Omega_{m_0},M,y_0),
\end{equation}
and $C_{stat+syst}$ is the covariance matrix that includes both statistical and systematic covariances. Here, $\Delta \vec{D}^T$ refers to the transpose of $\Delta \vec{D}$. The distance modulus $\mu = m - M$. Then, the total $\chi^2$ is obtained as,
\begin{equation}
    \chi^2_{total} = \chi^2_{OHD} + \chi^2_{SNIa}.
\end{equation}
\begin{figure}[h]
	\centering
	\includegraphics[width = 10cm]{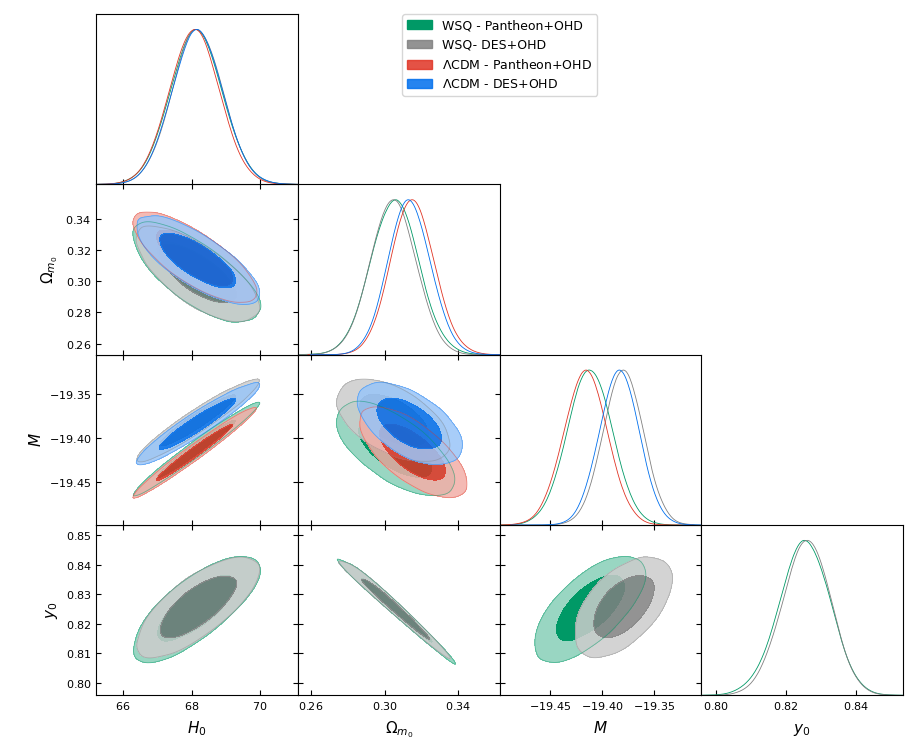}
	\caption{The 2D confidence contours for 68\% and 95\% probabilities and 1D posterior distribution of the model parameters using SNIa+OHD datasets for $\Lambda$CDM and the WSQ model. The chosen prior ranges are $40\le H_0\le100$, $0\le\Omega_{m_0}\le1$, $-20\le M \le -18$ and $0\le y_0 \le 1$.}
	\label{probdistribution}
\end{figure}

The parameter set ($H_0$, $\Omega_{m_0}$, $M$, $y_0$) that minimises the $\chi^2_{total} = -2\,\text{ln} (\mathcal{L})$, where $\mathcal{L}$ is the Gaussian likelihood, is the best-fit parameters. We utilize the information criteria such as Akaike Information Criterion (AIC) and Bayesian Information Criterion (BIC) for the model comparison. We use the definitions of AIC and BIC as,
\begin{align}
    &AIC = n\,\text{ln}\left(\frac{\chi^2}{n}\right) + 2k,\\
    &BIC = n\,\text{ln}\left(\frac{\chi^2}{n}\right) + k\,\text{ln}(n).
\end{align}
We used a well tested and open source PYTHON implementation of the affine-invariant ensemble sampler for MCMC proposed by Goodman and Weare called emcee for parameter inference. More details about emcee and its implementation can be found in Ref. \cite{emcee1,emcee2}. We also employed the multiprocessing module from PYTHON standard library and the numba module to reduce the computation time. The best-fit values of the model parameters are presented in Table \ref{modelparameters} and the marginal likelihood of the model parameters are presented in Fig. \ref{probdistribution}. The minimum $\chi^2$ obtained for the WSQ model with the SNIa + OHD dataset is lower than that obtained for $\Lambda$CDM. This suggests that the WSQ model is preffered over the $\Lambda$CDM according to the $\chi^2_{min}$ criteria. 
\begin{table}[ht]
    \centering
    \begin{tabular}{c@{\hskip 1.75 cm} c@{\hskip 1.75 cm} c}
        \hline \hline
         Data & Pantheon + OHD & DES + OHD\\
         \hline
         $H_0$ & $68.1041 \pm  0.7374$ & $68.2021 \pm 0.7374$\\
         $\Omega_{m_0}$ & $0.3059 \pm 0.0129$&$0.3040 \pm 0.0124$\\
         $M$ & $-19.4126 \pm 0.0212$&$-19.3789 \pm 0.0194$\\
         $y_0$ & $0.8251 \pm 0.0072$ & $0.8262 \pm  0.0069$\\
         \hline \hline
    \end{tabular}
    \caption{The best-fit model parameters of the WSQ model and its uncertainties within 1$\sigma$ confidence.}
    \label{modelparameters}
\end{table}
A similar trend is followed with the Akaike Information Criterion (AIC) value, however, the Bayesian Information Criterion (BIC) tends to penalize the extra parameter to a greater extent, so the $\Lambda$CDM is preferred over the WSQ model according to the $\Delta$BIC criteria. The information criteria we used to compare the fit of the models is given in Table \ref{aicbic}.
\begin{table}[ht]
    \centering
    \begin{tabular}{c@{\hskip 1.75 cm} c c c c}
    \hline \hline
         Dataset&Model &$\Delta\chi^2_{min}$ & $\Delta$AIC & $\Delta$BIC \\
         \hline
         \multirow{2}{4em} {Pantheon+OHD}&$\Lambda$CDM & 0 & 0 & 0\\
         & WSQ & 4.4514  &2.3579 & -3.1253\\
         \hline
         \multirow{2}{4em}{DES+OHD}&$\Lambda$CDM & 0 & 0 & 0\\
         & WSQ & 3.6300& 2.0642 & -3.4885 \\
         \hline \hline
    \end{tabular}
    \caption{The $\Delta\chi^2_{min}$, $\Delta AIC$ and $\Delta BIC$ obtained for the two models for the SNIa + OHD datasets.}
    \label{aicbic}
\end{table}
The estimated value for the Hubble constant, from both the datasets, is in close agreement with the Planck 2018 results \cite{refId0}. The current value of the dark energy density parameter is computed as $\Omega_{\phi_0} = 0.6941 \pm 0.0129$ with the Pantheon+OHD dataset and $\Omega_{\phi_0} = 0.6960 \pm 0.0124$ with the best fit values obtained for the DES+OHD dataset. With the estimated parameters, we analyse the background evolution of cosmological parameters within the WSQ background.
\section{Cosmographic Parameters}\label{evolution}
\begin{figure}[h]
    \centering
    \includegraphics[width = 10cm]{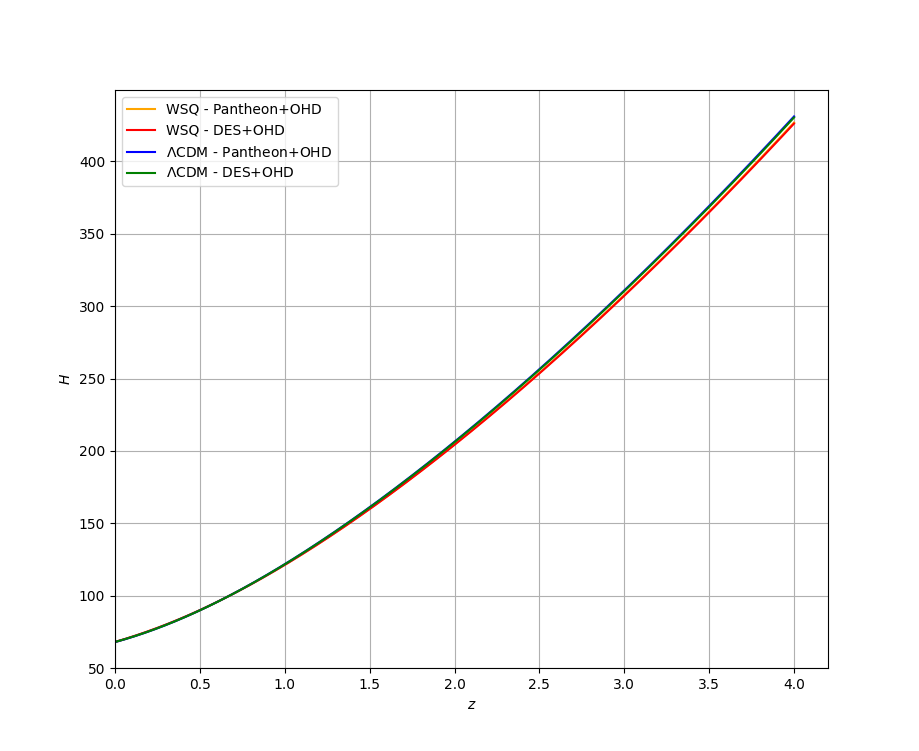}
    \caption{The evolution of Hubble parameter($H$) against redshift($z$) for the WSQ model and $\Lambda$CDM using the best-fit parameters for SNIa+OHD datasets.}
    \label{evohubble}
\end{figure}

This section deals with studying the evolution of various cosmographic parameters to understand the background evolution of the universe as explained by the WSQ model. One such parameter of profound  relevance is the Hubble parameter. The evolution of $H$ according to the WSQ model compared to that in the $\Lambda$CDM is shown in Fig. \ref{evohubble}. The evolution of the Hubble parameter is very similar to the $\Lambda$CDM model in the past and at present.

\begin{figure}[H]
    \centering
    \includegraphics[width = 10cm]{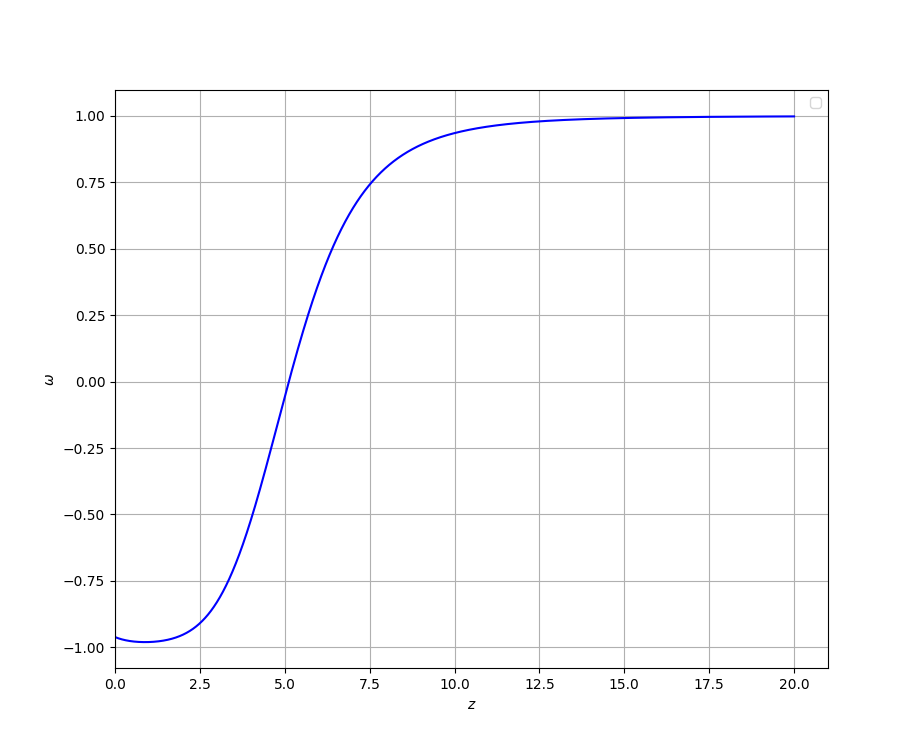}
    \caption{Evolution of the equation of state ($\omega$) against redshift($z$), plotted with the best-fit parameters obtained for SNIa+OHD datasets.}
    \label{eqnofstate}
\end{figure}

The evolution of the equation of state parameter $\omega$ of the scalar field $\phi$ is dynamical, while it is a constant in the $\Lambda$CDM with a value $= -1$ during the entire evolution of the Universe. The evolution of $\omega$ is presented in Fig. \ref{eqnofstate}. It is evident that for $z > 4.43$, the scalar field together with matter contributed to the deceleration. For  $z < 4.43$, the equation of state $-\frac{1}{3} < \omega < -1$, which shows the scalar field started contributing to the accelerated expansion, representing a freezing model of quintessence. The present value of $\omega$ is obtained as $-0.9617$. Furthermore, $\omega$ attains a value equal to 1 at larger redshifts, showing a stiff fluid nature of the scalar field at higher redshifts.

\begin{figure}[H]
	\centering
	\includegraphics[width = 10cm]{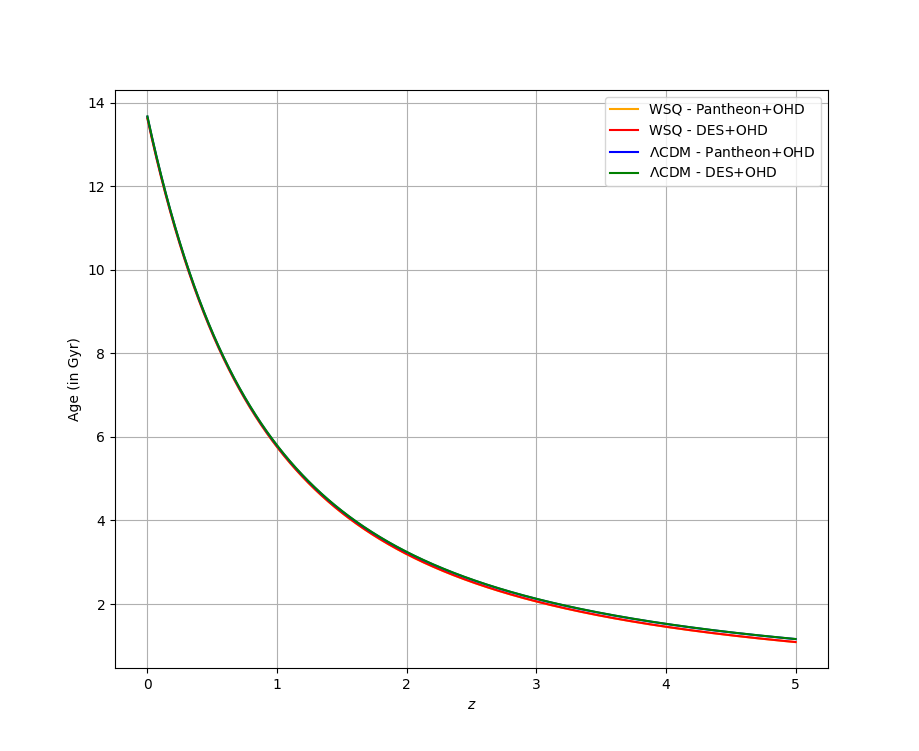}
	\caption{The age of the Universe against redshift($z$) is plotted for the WSQ model and $\Lambda$CDM using the best-fit parameters obtained for the SNIa + OHD datasets.}
	\label{ageofUni}
\end{figure}

The age of the Universe at a scale factor $a$ can be obtained from the definition of Hubble parameter $H = \dot{a}/a$ as,
\begin{equation}
    t_a - t_b = \int_0^a \frac{1}{aH(a)} da,
\end{equation}
where $t_a$ is the age of the Universe at scale factor $a$ and $t_b$ is the age at the moment of the big bang, which is assumed to be zero. The age of the Universe with respect to the redshift ($z$) is plotted in Fig. \ref{ageofUni}.

The present age of the universe is obtained at $z = 0$ and the computation results in an age of $13.6$ Gyr, which is slightly less than the age obtained for the $\Lambda$CDM from the Planck data. The age of the universe inferred from CMB Planck data with $\Lambda$CDM model \cite{refId0} is $13.8$ Gyr. The age of the Universe computed for the $\Lambda$CDM model with the same dataset is $13.7$ Gyr. 

\begin{figure}[h]
    \centering
    \includegraphics[width = 10cm]{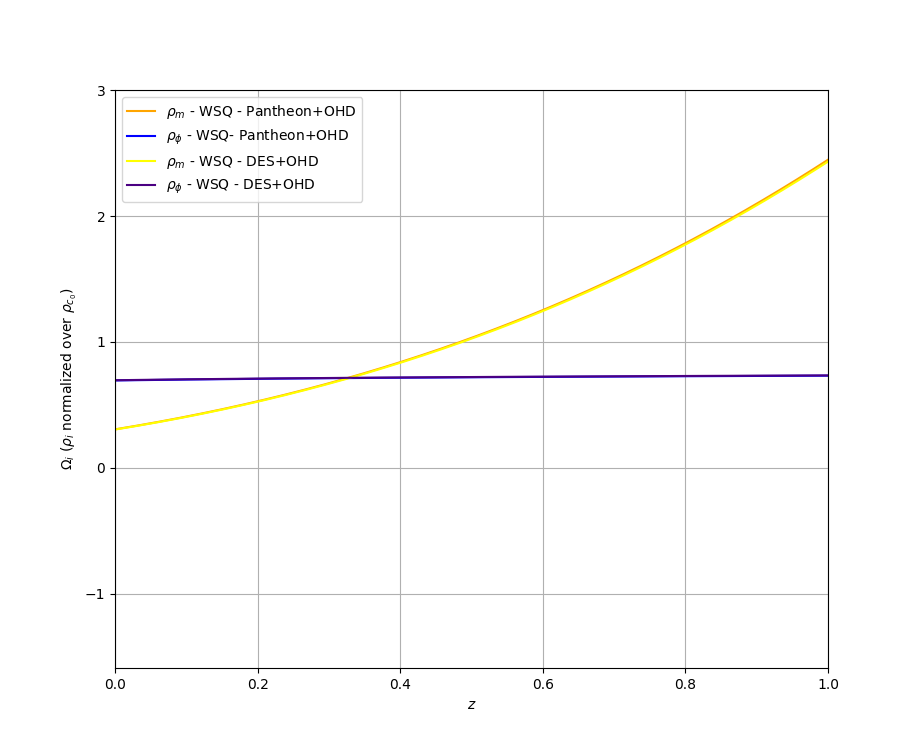}
    \caption{Evolution of matter and dark energy density in the WSQ model plotted with best-fit parameters obtained for the SNIa + OHD datasets.}
    \label{densityparam}
\end{figure}

The evolution of matter and dark energy density in the WSQ model is plotted in Fig. \ref{densityparam}. The matter energy density evolves exactly similar to the $\Lambda$CDM model, while the dark energy density or the energy density of the field slowly increases with redshift. It is clear that the dark energy dominates matter in late time, for redshift $z < 0.33$, which is in agreement with the predictions of standard $\Lambda$CDM.

\begin{figure}[ht]
     \centering
     \includegraphics[width = 10cm]{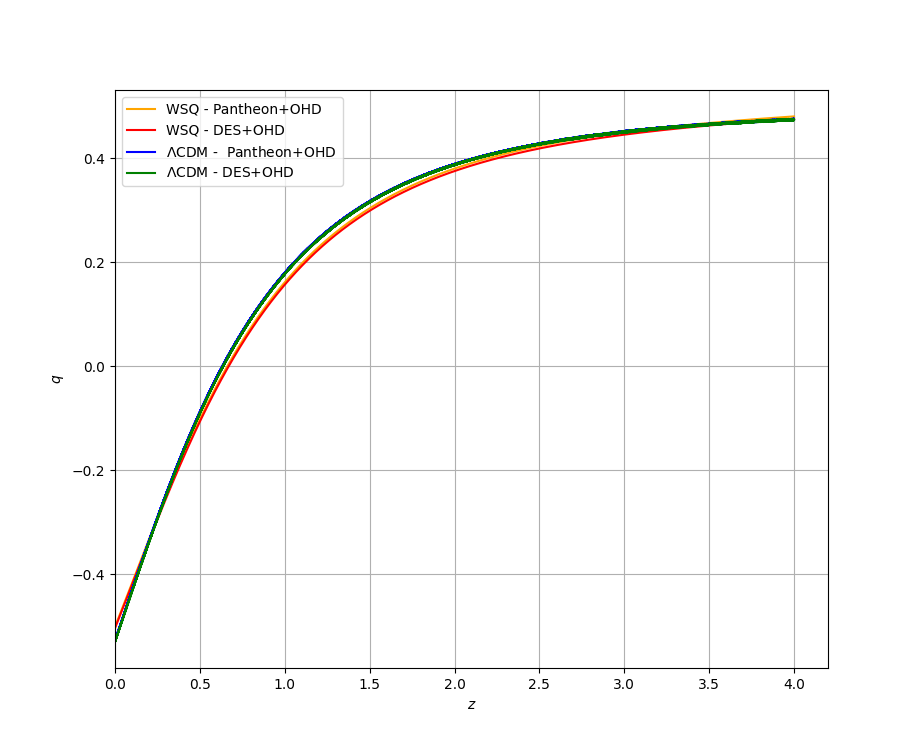}
     \caption{Evolution of deceleration parameter $q$ with redshift $z$ for the WSQ model and $\Lambda$CDM, plotted with best-fit parameters obtained for the SNIa+OHD datasets.}
     \label{decelplot}
 \end{figure}

The deceleration parameter $q$ serves as an indicator to the acceleration/deceleration of the Universe in the Friedmann-Lemaitre-Robertson-Walker (FLRW) background. It is expressed as,
 \begin{equation}\label{deceleq}
     q = -1 - \frac{\dot{H}}{H^2}.
 \end{equation}
In terms of redshift, we can express Eq. (\ref{deceleq}) as,
 \begin{equation}
     q = -1 + \frac{(1+z)}{H} \frac{dH}{dz}.
 \end{equation}
A negative value for $q$ would signify accelerated expansion of the Universe. In the WSQ model, we obtain the present deceleration parameter value of $q_0 = -0.501$, supporting the current accelerated expansion of the Universe. The redshift at which the Universe transitioned from decelerating to accelerating expansion is estimated to be $z_T = 0.662$. The WSQ model is successful in explaining the observed accelerated expansion of the Universe and the transition from decelerating to accelerating phase. The evolution of deceleration parameter with redshift is plotted in Fig. \ref{decelplot}.
 
 The statefinder pair $(r,s)$ is a diagnostic tool for studying dark energy models, initially introduced by Sahni et al. in Ref. \cite{rspair}, where $r$ is the jerk parameter and $s$ is a new parameter constructed from $r$ and the deceleration parameter $q$. The jerk parameter is defined as \cite{mattvisier2004},
 \begin{equation}
     r = \frac{1}{a}\frac{d^3a}{dt^3}\left[\frac{1}{a}\frac{da}{dt}\right]^{-3} = \frac{\dddot{a}}{aH^3}.
 \end{equation}
 We can represent $r$ in terms of the dimensionless Hubble parameter $h$ while introducing a new variable $N = \text{ln}(a)$, we obtain,
 \begin{equation}
     r = \frac{1}{2h^2}\frac{d^2h^2}{dN^2} + \frac{3}{2h^2}\frac{dh^2}{dN} + 1.
 \end{equation}
   \begin{figure}[H]
    \centering
    \includegraphics[width =10cm]{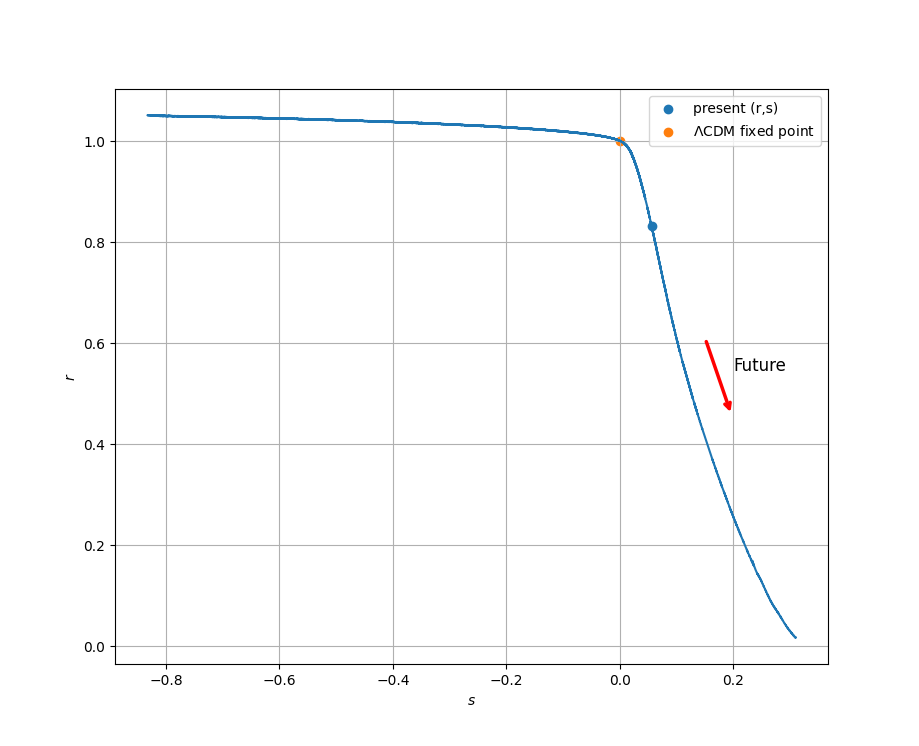}
    \caption{The statefinder trajectory is plotted using the best-fit parameters from the SNIa + OHD datasets. The fixed point of $\Lambda$CDM is also marked.}
    \label{rstraj}
\end{figure}
 The $s$ parameter is defined as ,
 \begin{equation}\label{sparam}
     s = \frac{r-1}{3(q - 1/2)}.
 \end{equation}
 On expressing $q$ presented in Eq. (\ref{deceleq}) in terms of $N$ and substituting in Eq. (\ref{sparam}), we obtain
 \begin{equation}
     s = -\frac{\frac{1}{2h^2}\frac{d^2h^2}{dN^2} + \frac{3}{2h^2}\frac{dh^2}{dN}}{\frac{3}{2h^2}\frac{dh^2}{dN} + \frac{9}{2}}.
 \end{equation}
 The statefinder trajectory distinguishes the behaviour of different types of dark energy models. The trajectory is plotted in Fig. \ref{rstraj}.
 The statefinder trajectory indicates the quintessence behavior of the field. The present value of the statefinder diagnostic pair is obtained as ($r,s$) = ($0.8309,0.0562$), different from the $\Lambda$CDM fixed point ($1,0$).

\begin{figure}[H]
	\centering
	\includegraphics[width=12cm]{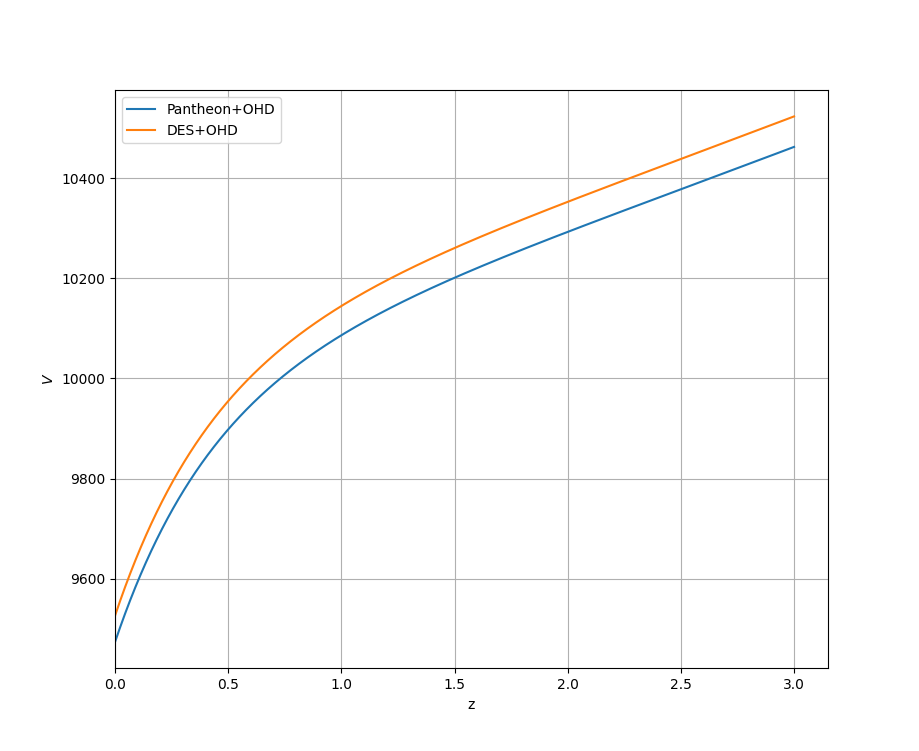}
	\caption{The evolution of the scalar field potential $V(\phi)$ plotted using the best-fit parameters from the SNIa + OHD dataset.}
	\label{potentialplot}
\end{figure}

\begin{figure}[H]
    \centering
    \includegraphics[width=10cm]{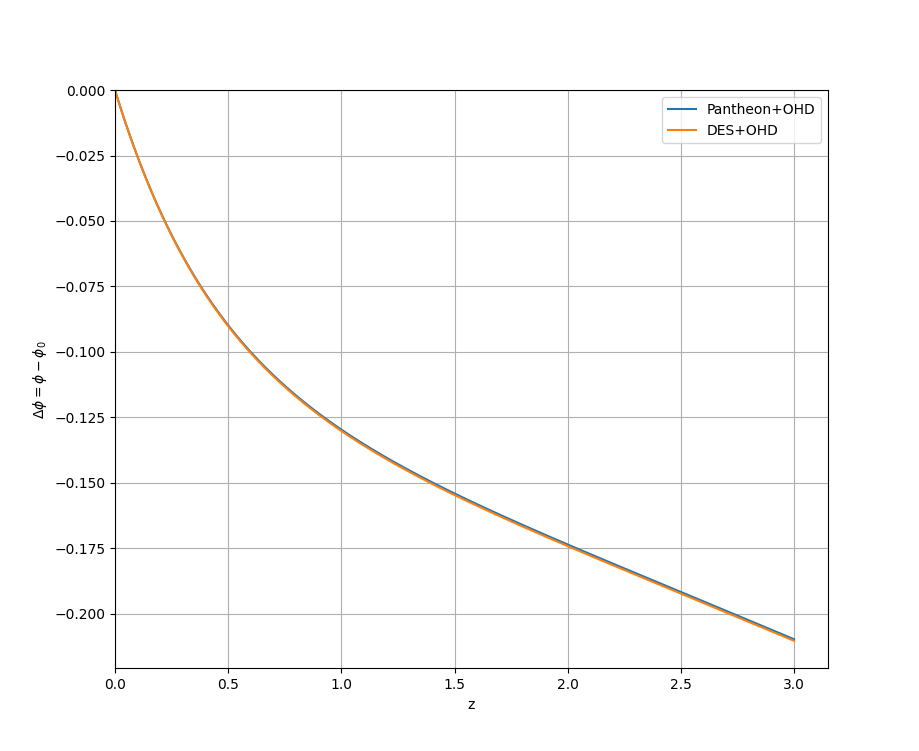}
    \caption{The qualitative evolution of the quantity $\Delta\phi = \phi - \phi_0$ plotted using the best-fit parameters from the SNIa + OHD dataset.}
    \label{fieldplot}
\end{figure}

We now study the physical properties of the scalar field, i.e., how the scalar field and its potential evolves with redshift. We present the qualitative evolution of the potential $V(\phi)$ as a function of redshift in Fig. \ref{potentialplot}. The potential is higher at earlier times as the scalar field rolls down the potential.
The qualitative evolution of the scalar field can be studied from the quantity $\Delta\phi(z) = \phi(z) - \phi_0$ obtained from Eq. (\ref{woodspotential}) and using the best fit parameters. This is presented in Fig. \ref{fieldplot}. The scalar field was initially lower than the present value and then increased at later times.
 \section{Dynamical System Analysis}\label{dynamicalanalysis}
\begin{figure}
    \centering
    \begin{tabular}{ll}
         \includegraphics[width = 7cm]{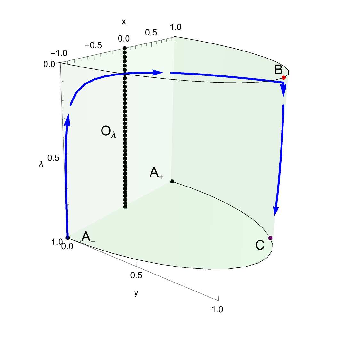}
         \includegraphics[width=7cm]{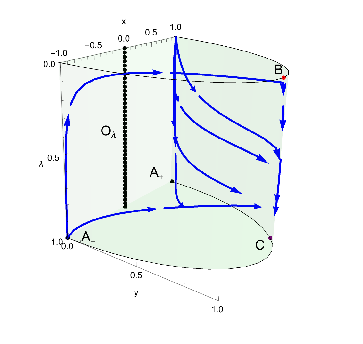}
    \end{tabular}
    \caption{The phase space trajectory for the WSQ model, depicting a single streamline on the left and multiple streamlines converging to the point $C$ on the right.}
    \label{phaseplot}
\end{figure}

In order to gain an understanding of the global behaviour any cosmological model, dynamical system analysis is particularly useful \cite{bahamonde2018dynamical,wainwright1997dynamical}, especially for studying its asymptotic behaviour. This requires a careful choice of dynamical variables to study the phase-space dynamics of the system.
\begin{equation}
   \label{dynamicalvariables} x = \frac{\dot{\phi}\kappa}{\sqrt{6}H},
    \hspace{0.2in} y = \frac{\kappa\sqrt{V(\phi)}}{\sqrt{3}H}.
\end{equation}
The dark energy or scalar field density is then defined as,
\begin{equation}
    \Omega_\phi = \frac{\kappa^2\rho_\phi}{3H^2} = x^2 + y^2.
\end{equation}
The effective equation of state of the background cosmological fluid would be,
\begin{equation}
    \omega_{eff} = \frac{P_\phi + P_m}{\rho_\phi + \rho_m} = x^2 - y^2
\end{equation}
From Eq. (\ref{mot1}) and (\ref{mot2}), we get the relation,
\begin{equation}
    \frac{\dot{H}}{H^2} = -\frac{3}{2}(1 + x^2 - y^2).
\end{equation}
Using this equation and Eq. (\ref{dynamicalvariables}), we differentiate these variables with respect to $N = \text{ln}(a)$, to obtain a system of autonomous differential equations. The dynamical variables $x$ and $y$ cannot close the system of autonomous equations without considering another dynamical variable $\lambda$, which is defined as,
\begin{equation}\label{lambdadef}
    \lambda = -\frac{V_{,\phi}}{\kappa V(\phi)} = \left(1 - \frac{V(\phi)}{2V_0}\right).
\end{equation}
The system of autonomous differential equations are then given by,
\begin{align}
    \label{autonomous1}&\frac{dx}{dN} = -3x + \frac{\sqrt{6}}{2}\lambda y^2 + \frac{3}{2}x(1 + x^2 - y^2) \\
   \label{autonomous2}&\frac{dy}{dN} = -\frac{\sqrt{6}}{2}\lambda xy + \frac{3}{2}y(1 + x^2 - y^2)\\
    \label{autonomous3}&\frac{d\lambda}{dN} = \sqrt{6}x\lambda(1 - \lambda).
\end{align}
Before studying the dynamics, we need to make sure that the phase space of this system is compact. The variables $x$ and $y$ are constraint by Friedmann equation (\ref{mot1}). The variable $\lambda$ is also compact, from the fact that the minimum value of $V(\phi)$ is zero and the maximum value is effectively $2V_0$. Thus the phase space of this system is represented by the positive-$y$ half cylinder from $\lambda = 0$ to $\lambda = 1$. The phase space portrait for this system is given in Fig. \ref{phaseplot}.
The fixed points of the system can be obtained by setting Eq. (\ref{autonomous1}), (\ref{autonomous2}) and (\ref{autonomous3}) to zero. These critical points and their properties are outlined in Table \ref{criticaltable}.
\begin{table}[H]
    \centering
    \begin{tabular}{|c|c|c|c|c|c|c|c|}
    \hline
          Point & $x_c$ &$y_c$ &$\lambda_c$&Accel.&$\Omega_\phi$&$\omega_{eff}$&$\omega_\phi$\\[0.25 cm]
         \hline
         $O_\lambda$&0&0&Any&No&0&0&-\\[0.25 cm]
         $A_\pm$&$\pm1$&0&1&No&1&1&1\\[0.25 cm]
         $B$&0&1&0&Yes&1&-1&-1\\
         $C$&$\frac{1}{\sqrt{6}}$&$\sqrt{\frac{5}{6}}$&1&Yes&1&-$\frac{2}{3}$&-$\frac{2}{3}$\\[0.25 cm]
         \hline
    \end{tabular}
    \caption{Critical Points}
    \label{criticaltable}
\end{table}
It is to be noted that the dynamical variables $x$, $y$ and $\lambda$ are constrained to lie in the ranges $-1 \le x \le 1$, $0 \le y \le 1$, and $0 \le \lambda \le 1$ as per their definition (Eq. \ref{dynamicalvariables}) and Eq. (\ref{mot1}). So critical points that lie outside this range have not been considered. Additionally, when $y \rightarrow 0$, $\lambda \rightarrow 1 $ according to Eq. (\ref{lambdadef}), and critical points which do not satisfy this constraint have also been not considered. In order to study the stability of these fixed points $(x,y,\lambda) = (x_c,y_c,\lambda_c)$, we consider linear perturbations about these points $\delta x,\delta y$ and $\delta\lambda$ such that they satisfy,
\begin{equation}
\label{perturbationequation}
    \frac{d}{dN} \begin{pmatrix}
        \delta x\\
        \delta y\\
        \delta y
    \end{pmatrix}
    = J\begin{pmatrix}
        \delta x\\
        \delta y\\
        \delta y
    \end{pmatrix}.
\end{equation}
where the Jacobian matrix $J$ is given by,
\begin{equation}
    J = \begin{pmatrix}
        \frac{\partial f}{\partial x} & \frac{\partial f}{\partial y} & \frac{\partial f}{\partial \lambda}\\[9pt]
         \frac{\partial g}{\partial x} & \frac{\partial g}{\partial y} & \frac{\partial g}{\partial \lambda}\\[9pt]
          \frac{\partial h}{\partial x} & \frac{\partial h}{\partial y} & \frac{\partial h}{\partial \lambda}
    \end{pmatrix}_{x = x_c,y = y_c,\lambda = \lambda_c},
\end{equation}
where $f(x,y,\lambda)$, $g(x,y,\lambda)$ and $h(x,y,\lambda)$ are the RHS of Eq. (\ref{autonomous1}),(\ref{autonomous2}),(\ref{autonomous3}) respectively. The stability of the critical points is determined by the eigenvalues of this Jacobian matrix at these critical points. If all eigenvalues are negative, the critical point is a future attractor, which is stable. If all the eigenvalues are positive, the corresponding point is a past attractor and is unstable and if some eigenvalues are positive and some negative, the critical point is a saddle point. If the eigenvalues are complex, the critical points would be stable spiral or unstable spiral depending on the real parts of the eigenvalues. These points with eigenvalues having non-zero real part are called $hyperbolic$ points \cite{bohmer2017dynamical}. The linear stability theory fails if any of the eigenvalues is zero and other methods have to be employed to determine the stability of the system at this $non-hyperbolic$ critical point. The eigenvalues of the critical points presented in Table \ref{criticaltable} and their stability are given in Table \ref{stabilitytable},
\begin{table}[H]
    \centering
    \begin{tabular}{|c|c|c|c|}
    \hline
    Point&Eigenvalues&Nature of critical point&Stability\\[0.25 cm]
    \hline& & &\\
    $O_\lambda$&$\left(-\frac{3}{2},\frac{3}{2},0\right)$&Non-hyperbolic& Saddle\\[0.25 cm]
    $A_\pm$&$\left(3,\pm\sqrt{6},\frac{1}{2}(6 \pm \sqrt{6})\right)$&Hyperbolic&Unstable/Saddle\\[0.25 cm]
    $B$& $\left(-3,-3,0\right)$&Non-hyperbolic&Saddle\\[0.25 cm]
    $C$&$\left(-\frac{5}{2},-2,-1\right)$&Hyperbolic&Stable\\[0.25 cm]
    \hline
    \end{tabular}
    \caption{Critical Points and their eigenvalues}
    \label{stabilitytable}
\end{table}
For non-hyperbolic critical points, there are usually two approaches to study their stability. One is using Lyapunov functions\cite{brauer1989qualitative,TiagoCCharters_2001}, and the other is the Centre Manifold theorem. Here we have employed the centre manifold theorem to analyze the stability of the non-hyperbolic point $B$.
\subsection{Analysis of critical points}
\begin{itemize}
    \item \textbf{Points $O_\lambda$} : This set of points correspond to the $\lambda$-axis, i.e., the $\lambda$-axis acts as a critical line. On this line, the Universe is matter-dominated, with the scalar field energy density (and pressure) vanishing at this critical line, leaving $\omega_\phi$ undetermined and the effective equation of state (EoS) $\omega_{eff}$ equal to the matter equation of state, i.e., zero. This gives a non-accelerating solution because of the absence of the scalar field. From the eigenvalues of the critical points, one eigenvalue vanishes and hence this set of points is non-hyperbolic and linear stability is inapplicable. But from the remaining two eigenvalues, one is positive and the other is negative. Thus, the critical points $O_\lambda$ is asymptotically unstable and thus will act as a saddle line.
    \item \textbf{Points $A_\pm$} : These two points $(\pm 1,0,1)$correspond to scalar field dominated solutions where the kinetic component of the scalar field dominates. The Universe is completely dominated by the scalar field and results in an effective EoS $\omega_{eff} = 1$, which characterizes a stiff fluid dominated solution. Both points are associated with non-accelerated expansion of the Universe. The eigenvalues corresponding to these points suggest that they are unstable, with $A_+$ being an unstable point (a past attractor) and $A_-$ being a saddle point.
    \item \textbf{Point $B$} : This point $(0,1,0)$ corresponds to a cosmological constant-like dominated solution where the Universe undergoes a de Sitter accelerated expansion, with the effective EoS $\omega_{eff} = -1$ and the scalar field completely dominating the Universe. This point represents the scalar field potential energy dominated solution with vanishing kinetic energy. From Table \ref{stabilitytable}, it is evident that this point is non-hyperbolic, because of the vanishing eigenvalue. Hence, linear stability is not applicable and to determine whether this point could act as a stable future attractor, we apply Center manifold theorem. Through this approach (Sec. \ref{centremanifoldsection}), we find that this critical point is asymptotically unstable, indicating that it functions as a saddle point in the dynamics of the system.
    \item \textbf{Point $C$} : This point $\left(\frac{1}{\sqrt{6}},\sqrt{\frac{5}{6}},1\right)$ also represents a solution where the Universe is entirely dominated by the scalar field as the scalar field energy density $\Omega_\phi = 1$. Unlike Point $B$, this point has non-zero contributions from both the kinetic and potential energy components of the scalar field. This solution gives an effective EoS $\omega_{eff} = -\frac{2}{3}$, resulting in accelerated expansion of the Universe. Since the eigenvalues are non-zero, linear stability theory can be applied to this point. With all three eigenvalues being negative, this indicates that the point is a stable future attractor.
\end{itemize}
\subsection{Centre Manifold Approach to the Point $B$}\label{centremanifoldsection}
The critical point $B$ $(0,1,0)$ has eigenvalues $(-3,-3,0)$ and linear stability theorem cannot be applied to study the stability of this point. In order to apply the centre manifold theorem, we first shift the critical point to the origin, by transforming the dynamical variables in Eq. (\ref{dynamicalvariables}) into a new set of variables,
\begin{equation}
    X = x, \hspace{0.2 in} Y = y-1,\hspace{0.2 in} Z = \lambda.
\end{equation}
Then the derivatives with respect to $N = \text{ln}(a)$ for these new variables are given by,
\begin{align}
    \label{newderiv1}&\frac{dX}{dN}=-3X +\frac{\sqrt{6}}{2}Z(Y+1)^2+ \frac{3}{2}X(1+X^2-(Y+1)^2)\\
    \label{newderiv2}&\frac{dY}{dN}=-\frac{\sqrt{6}}{2}ZX(Y+1)+\frac{3}{2}(Y+1)(1 +X^2+(Y+1)^2)\\
    \label{newderiv3}&\frac{dZ}{dN}=\sqrt{6}XZ(1-Z).
\end{align}
Now we find the Jacobian for this set of autonomous equations at the origin (0,0,0). We obtain the Jacobian as
\begin{equation}
    J\big|_{X=0,Y=0,Z=0}  = \begin{pmatrix}
        -3&0&\frac{\sqrt{6}}{2}\\
        0&-3&0\\
        0&0&0
    \end{pmatrix}.
\end{equation}
Now, we introduce a new set of variables such that,
\begin{equation}\label{transformationeqn}
    \begin{pmatrix}
        U\\
        V\\
        W
    \end{pmatrix} = S^{-1}\begin{pmatrix}
        X\\
        Y\\
        Z
    \end{pmatrix},
\end{equation}
where $S$ is the matrix formed by the eigenvectors of the Jacobian matrix $J$. We obtain the matrix $S$ as,
\begin{equation}
    S = \begin{pmatrix}
        1&0&\frac{1}{\sqrt{6}}\\
        0&1&0\\
        0&0&1
    \end{pmatrix}
\end{equation}
Then, from  Eq. (\ref{transformationeqn}), we get the new set of variables as,
\begin{equation}
    U = X - \frac{Z}{\sqrt{6}}, \hspace{0.2 in}
    V = Y, \hspace{0.2 in}
    W = Z.
\end{equation}
Under this transformation, Eq.(\ref{newderiv1}),(\ref{newderiv2}) and (\ref{newderiv3})can be rewritten as,
\begin{flalign}
\label{dUdN}
\begin{split}
    \frac{dU}{dN} =&-3\left(U + \frac{W}{\sqrt{6}}\right) +  \frac{\sqrt{6}}{2}W(V+1)^2\\& +\frac{3}{2}\left(U +\frac{W}{\sqrt{6}}\right)\left[1 - \left(U + \frac{W}{\sqrt{6}}\right)^2 + (V+1)^2\right]
    - \left(U + \frac{W}{\sqrt{6}}\right)W(1 - W),
\end{split}\\
    \label{dVdN}\frac{dV}{dN} =&-\frac{\sqrt{6}}{2}W\left(U + \frac{W}{\sqrt{6}}\right)(V+1) + \frac{3}{2}(V+1)\left[1 - \left(U + \frac{W}{\sqrt{6}}\right)^2 + (V+1)^2\right],\\
    \label{dWdN}\frac{dW}{dN} =& \sqrt{6}\left(U + \frac{W}{\sqrt{6}}\right)W(1-W).
\end{flalign}
According to the centre manifold theory, these equations can be represented in the form,
\begin{align}
    &\frac{dU}{dN} = AU + F(U,V,W)\\
    &\frac{dV}{dN} = BV + G(U,V,W)\\
    &\frac{dW}{dN} = CW + H(U,V,W)
\end{align}
where we obtain $A = -3$, $B = -3$ and $C = 0$. Then, by centre manifold theorem, there exists a centre manifold such that the dynamics of the system reduced to the centre manifold is given by,
\begin{equation}
    \label{centremanifoldeq}\frac{dW}{dN} = CW + H(h_1(W),h_2(W),W),
\end{equation}
for sufficiently small $W$. Here we have assumed $U = h_1(W)$ and $V = h_2(W)$. This should satisfy the quasilinear partial differential equations,
\begin{align}
\label{quasi1}&\frac{dh_1(W)}{W}\left(CW + H(h_1(W),h_2(W),W)\right) - \left(Ah_1(W) + F(h_1(W),h_2(W),W)\right) = 0\\
\label{quasi2}&\frac{dh_2(W)}{W}\left(CW + H(h_1(W),h_2(W),W)\right) - \left(Bh_2(W) + G(h_1(W),h_2(W),W)\right) = 0
\end{align}
Since an exact knowledge of the centre manifold is not necessary, it suffices to approximate the centre manifold\cite{haasdonk2021kernel} and hence it is customary to assume the expansions,
\begin{align}
    h_1(W) = a_2W^2 + a_3W^3 + \mathcal{O}(W^4)\\
    h_2(W) = b_2W^2 + b_3W^3 + \mathcal{O}(W^4).
\end{align}
Substituting this into Eq. (\ref{quasi1}) and (\ref{quasi2}), we can equate the coefficients of each power of $W$ equal to zero to obtain the values of the coefficients $a_2$, $a_3$, $b_2$ and $b_3$. We obtain $a_2 = -\dfrac{1}{3\sqrt{6}}$, $a_3 = \dfrac{2}{3\sqrt{6}}$, $b_2 = -\dfrac{1}{12}$ and $b_3 = \dfrac{1}{18}$. Thus Eq. (\ref{centremanifoldeq}) becomes,
\begin{equation}\label{centremanifoldapprox}
\begin{split}
        \frac{dW}{dN} &= W^2 - W^3 + \sqrt{6}\left(a_2W^2 + a_3W^3\right)W - \sqrt{6}\left(a_2W^2 + a_3W^3\right)W^2\\
        &=W^2 - \frac{4}{3}W^3 + W^4 + \mathcal{O}(W^5)
\end{split}
\end{equation}
From Eq. (\ref{centremanifoldapprox}), it can be concluded that for sufficiently small W, the system is unstable as the lowest order term has even parity. Hence the critical point $(0,1,0)$ is unstable and acts as a saddle point.

To conclude the results of the dynamical system analysis, we obtained one stable critical point (Point $C$), two saddle points and a saddle line, and an unstable critical point. Thus, the Universe is asymptotically stable within the WSQ model and attains an accelerated expansion. Since the WSQ model fits observation and predicts parameter values consistent with that of the standard model, and is dynamically stable, this model may be considered as a potential alternative to the standard model.

\section{Conclusion}\label{conclusion}
In the presented work, we examined the scalar field model with Woods-Saxon like potential. In an earlier work, Oikonomou and Chatzarakis demonstrated that early inflation can be successfully realized using a scalar field model with a similar potential. In this work, we study the implications of such a scalar field in the late Universe in producing the observed acceleration. 

We tested the model with the latest observational data using the data combination SNIa + OHD and adopting Bayesian Inference. Our analysis shows that the model is consistent with the observational data and the obtained values of parameters are also consistent with that obtained with $\Lambda$CDM for the same dataset. The obtained best-fit values of model parameters are $H_0 = 68.10 \pm 0.74$, $\Omega_{m_0} = 0.31 \pm 0.01$ and $y_0 = 0.83 \pm 0.01 $, which is in close agreement with $\Lambda$CDM predictions. The WSQ model is preferred over the $\Lambda$CDM according to $\chi^2_{min}$ and AIC, while according to the BIC criteria, the $\Lambda$CDM is slightly preferred over the WSQ model, because of the extra parameter.

The study of evolution of different cosmological parameters also revealed the consistency of the WSQ model in explaining the late-time background evolution of the Universe. The evolution of $\omega$ shows a freezing model of quintessence. Age of the Universe is estimated to be $13.6$ Gyr, which is slightly less than that predicted by the $\Lambda$CDM. The present value of the deceleration parameter obtained with the WSQ model is negative, showing that the model predicts present day accelerated expansion. The Universe transitioned from decelerating to accelerating is at a redshift $z_T = 0.662$. The statefinder trajectory reveals the quintessence behaviour of the field.

We performed dynamical system analysis to test the stability of the Universe in the WSW model. Studying the evolution of the dynamical variables $\kappa\dot{\phi}/\sqrt{6}H$, $\kappa\sqrt{V(\phi)}/\sqrt{3}H$ and $\lambda = (1 - V(\phi)/2V_0)$ from their autonomous coupled differential equations, we obtained multiple critical points, of which only one was a stable attractor according to linear stability theorem. One non-hyperbolic critical point was studied using the centre manifold approach and was found to be unstable. In conclusion, the WSQ model does allow for a future attractor and the Universe is stable within the WSQ model.

In summary, the model is consistent with the observations to good extent and is dynamically stable. Thus, this model can be considered as a potential alternative to the $\Lambda$CDM to explain the late phase acceleration of the Universe. This complements the findings of \cite{woods-saxoninflation}, which studied the early phase inflation and reheating with a scalar field model with a similar Woods-Saxon potential, reinforcing the possibility of such a component driving accelerated expansion in the late Universe.

The Hubble tension is one of the most important challenges faced by modern cosmology. Our analysis shows that the Hubble parameter value obtained for the WSQ model is close to the CMB predicted value assuming the standard $\Lambda$CDM, while there exists a 3.9$\sigma$ tension with the Hubble constant value obtained by the SH0ES collaboration \cite{riess2022comprehensive}.
In this work, we considered a quintessence which does not couple with ordinary matter. There are attempts in literature to alleviate the Hubble tension by introducing interaction between dark matter and the field \cite{Barros_2019}. Hence, we extend this study by introducing non-minimal coupling between matter and field as an attempt to alleviate the Hubble tension.


\acknowledgments

The authors of the manuscript are thankful to the Indian Institute of Technology Kanpur for providing the Param Sanganak high-performance computation facility for faster execution of the Python program. One of the authors, Sarath Nelleri, is thankful to the Indian Institute of Technology Kanpur for providing the Institute Postdoctoral Fellowship. The author Sreerag Radhakrishnan is thankful to Mrs. Sanitha A for fruitful discussions on the topic.





\providecommand{\href}[2]{#2}\begingroup\raggedright\endgroup

\end{document}